\documentclass[a4paper,11pt]{article}
\pdfoutput=1 

\usepackage{jcappub} 

\usepackage[T1]{fontenc} 

\usepackage{mathtools}
\usepackage{ulem}
\usepackage{comment}
\usepackage{siunitx}

\title{\boldmath Simulations of systematic effects arising from cosmic rays in the LiteBIRD space telescope, and effects on the measurements of CMB $B$-modes}

\author[a,b,1]{S. L. Stever,\note{Corresponding author.}}
\author[b]{T. Ghigna}
\author[c,d]{M. Tominaga}
\author[e]{G. Puglisi}
\author[d]{M. Tsujimoto}
\author[f]{M. Zeccoli Marazzini}
\author[f]{M. Baratto}
\author[f]{M. Tomasi}
\author[g]{Y. Minami}
\author[h]{S. Sugiyama}
\author[i]{A. Kato}
\author[b]{T. Matsumura}
\author[a]{H. Ishino}
\author[i]{G. Patanchon}
\author[j]{M. Hazumi}

\affiliation[a]{Okayama University, 3-1-1 Tsushimanaka, Kita-ku, Okayama, 700-8530 Japan}
\affiliation[b]{Kavli Institute for the Physics and Mathematics of the Universe (WPI),The University of Tokyo Institutes for Advanced Study, The University of Tokyo, Kashiwa, Chiba, 277-8583 Japan}
\affiliation[c]{Department of Astronomy, The University of Tokyo, Bunkyo-ku, Tokyo, 113-8654 Japan}
\affiliation[d]{Japan Aerospace Exploration Agency, Institute of Space and Astronautical Science, Sagamihara, Kanagawa, 252-5210 Japan}
\affiliation[e]{University of California - Berkeley, Berkeley, CA, United States}
\affiliation[f]{Università degli Studi di Milano, 20122, Milano, Italy}
\affiliation[g]{Research Center for Nuclear Physics, Osaka University, Ibaraki, Osaka, 567-0047 Japan}
\affiliation[h]{Saitama University, 255 Shimookubo, Sakura-ku, Saitama, 338-8570 Japan}
\affiliation[i]{Université de Paris, CNRS, Astroparticule et Cosmologie, F-75006 Paris, France}
\affiliation[j]{KEK - High Energy Accelerator Research Organization, Tsukuba, Ibaraki, 305-0801 Japan}

\emailAdd{sstever@okayama-u.ac.jp}
\emailAdd{tommaso.ghigna@ipmu.jp}
\emailAdd{tominaga@ac.jaxa.jp }
\emailAdd{tsujimot@astro.isas.jaxa.jp}
\emailAdd{yminami@rcnp.osaka-u.ac.jp}
\emailAdd{sugiyama@heal.phy.saitama-u.ac.jp}
\emailAdd{akikato@post.kek.jp}
\emailAdd{tomotake.matsumura@ipmu.jp}
\emailAdd{scishino@s.okayama-u.ac.jp}
\emailAdd{masashi.hazumi@kek.jp}

\abstract{Systematic effects arising from cosmic rays have been shown to be a significant threat to space telescopes using high-sensitivity bolometers. The LiteBIRD space mission aims to measure the polarised Cosmic Microwave Background with unprecedented sensitivity, but its positioning in space will also render it susceptible to cosmic ray effects. We present an end-to-end simulator for evaluating the expected scale of cosmic ray effect on the LiteBIRD space mission, which we demonstrate on a subset of detectors on the 166 GHz band of the Low Frequency Telescope. The simulator couples the expected proton flux at L2 with a model of the thermal response of the LFT focal plane and the electrothermal response of its superconducting detectors, producing time-ordered data which is projected into simulated sky maps and subsequent angular power spectra.}

\begin{document}
\maketitle
\flushbottom

\section{Introduction}
\label{sec:intro}
LiteBIRD is a next-generation cosmological full-sky survey led by the Japan Aerospace Exploration Agency (JAXA) with the goal of measuring polarised primordial $B$-modes.
LiteBIRD is planned for launch in the late 2020s, after which it will join an orbit at the second Earth-Sun Lagrangian point (L2)~\cite{sugai2020updated}.
The ability of LiteBIRD to achieve its scientific goals is greatly amplified by its space-borne positioning, due to a lack of atmospheric interference with the Cosmic Microwave Background (CMB) signal and the ability to measure the sky at large angular scales.
The tensor-to-scalar ratio $r$ determines the power of the $B$-mode signal, and LiteBIRD's design goal includes achieving a sensitivity $\delta r < 0.001$.
However, the goal of measuring the entire sky with such sensitivity requires precise characterisation and control over systematic effects, particularly cosmic rays.

The last major Cosmology space mission was the ESA-led Planck space telescope~\cite{tauber2004planck},
which consisted of both a High Frequency Instrument (HFI) and a Low Frequency Instrument (LFI).
Planck HFI and LiteBIRD are similar in two points: (\textit{i}.) the L2 orbit at which measurements are taken, and (\textit{ii}.) use of highly-sensitive low-temperature bolometers.
As bolometers are sensitive to any change in temperature, they are also vulnerable to the thermal fluctuations induced by cosmic ray energy deposition.

\subsection{Overview of cosmic rays as a systematic effect}
At L2, there are two populations of cosmic rays: 
Galactic Cosmic Rays (GCRs) and solar cosmic rays.
GCRs originate from outside the solar system, from within or beyond the local Galaxy, and have a peak energy distribution range of 0.1--\SI{1}{GeV},
although energies up to 3--\SI{108}{TeV} have been observed~\cite{aharonian1999time}.
The ratio between these two populations is strongly affected by the behaviour of solar magnetic fields~\cite{gleeson1968solar} which oscillate in an 11 year cycle, consisting of 11 years of minimal solar activity followed by 11 years of maximum solar activity.
During so-called solar maxima, the presence of the solar magnetic field is at its strongest, and this magnetic field significantly attenuates the flux of GCRs at L2. Solar maxima also produce the most solar flares which, despite being transient events, have the ability to saturate larger areas of the telescope focal plane and render data unusable during specific periods of time.
Conversely, during solar minima, the GCR flux is strongest due to being the least inhibited by solar magnetic fields. This results in a higher constant flux of protons impacting the spacecraft.

Planck was launched during the 2009 solar minimum, which was also the year for the highest cosmic ray flux since recording began in 1964~\cite{usoskin2001dependence}.
Subsequently, the first data from Planck HFI showed a higher-than-expected rate of cosmic ray `glitches' and an associated variability of the thermal load of the cold plate of the instrument.
This rate of cosmic ray impacts on the HFI detectors, as well as the unexpected sensitivity of these detectors to these impacts, would have resulted in $90\%$ of the data being unusable due to long glitch tails if great effort had not been made to fit to and remove the CR signal~\cite{ade2014planck,catalano2014characterization,catalano2014impact}.
Whilst the majority of the cosmic ray systematic effects were eliminated in the final data analysis, the experimental characterisation of radiation interactions on the HFI detectors was largely post-facto.

\begin{figure}[tbp]
\centering 
\includegraphics[width=0.9\textwidth]{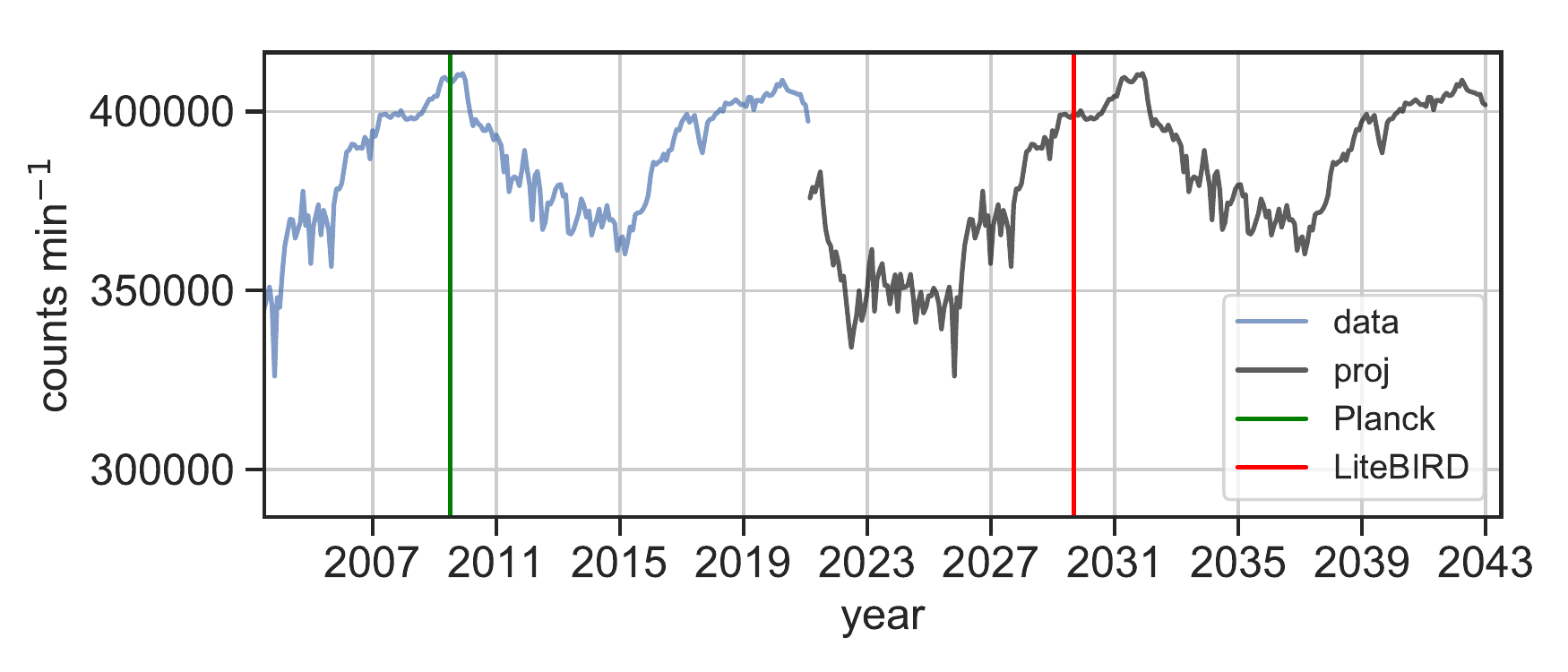}
\caption{\label{fig:crfluxoulu} Oulu neutron observatory data up to the present day (blue), the projected data for the next 22 years (black), the observation time of Planck HFI (green line), and the observation time of LiteBIRD (red line).}
\end{figure}

Any space mission employing high-sensitivity bolometers or calorimeters will grapple with the presence of cosmic rays, and solutions to the cosmic ray problem are an open topic of study for CMB, infrared, and X-ray missions (e.g. Athena X-IFU~\cite{stever2019characterisation}, JWST, WFIRST, etc.).
Due to common attributes of instrument hardware in these missions, progress on cosmic ray issues is applicable both within and outside Cosmology.

The observation period of LiteBIRD will take place during the first solar minimum since the observations of Planck.
Using data from the Oulu neutron observatory, which has been monitoring neutron flux on Earth since 1964, we can compare the relative levels of cosmic rays over the complete solar cycle.
By projecting the past 22 years of data into the next 22 years, we can predict that LiteBIRD will be coming online during the years of minimal solar activity (and peak GCR flux).
The abridged Oulu data, corrected for baryometric pressure, the projected future neutron count levels, and the observation times of Planck and LiteBIRD are shown in figure~\ref{fig:crfluxoulu}. This shows that cosmic ray effects are of particular importance for the success of the LiteBIRD mission.

Especially in the age of precision Cosmology, precise understanding and control over systematic effects is paramount.
In LiteBIRD, studies have already been conducted to study cosmic ray mitigation methods in its highly-sensitive detectors from a hardware perspective~\cite{minami2020irradiation}.
At present, we hope to couple these experimental studies with simulations, and thus present the current status of an end-to-end simulator developed for this purpose, using which we show first results on the \SI{119}{GHz} band of the Low Frequency Telescope (LFT) on LiteBIRD as a proof-of-concept.

\subsection{Physical description of cosmic ray effects in LiteBIRD}
At L2, GCRs (primarily protons) will impact the spacecraft and impart a portion of their energy as they traverse through it.
Upon impacting with the aluminium frame and components of the spacecraft hull and hardware, protons with enough energy to cross through the spacecraft will produce showers of secondary particles (largely electrons) which will impact the focal plane and detector arrays.
As the electrons deposit their energy into the cryogenic detector wafer, energy will propagate as heat and the wafer temperature will rise.

Although the specific mechanisms by which Planck HFI was susceptible to cosmic ray flux have been widely elucidated in studies subsequent to its 2009 flight \cite{catalano2014characterization}\cite{catalano2014impact}, the design of the LiteBIRD detectors and focal plane are significantly different, thus requiring LiteBIRD-specific predictions and treatment. 
The largest difference is the presence of many detectors on a single silicon wafer, rather than isolated absorbers on a per-detector basis like Planck HFI. 
This difference emphasises coincidence effects over many detectors at the same time arising from energy propagation across the wafer surface, which is an issue typical for large arrays with coupling to a common thermal surface.
Furthermore, the electrothermal response of LiteBIRD's TES detectors is naturally different to that of Planck HFI's semiconducting spider web bolometers, and there are also significant differences in the readout technology, especially in the multiplexing, filtering and decimation stages.

As our first demonstration of the simulator focuses on the Low Frequency Telescope (LFT) of LiteBIRD, we have simulated its detectors and detector wafer.
The detectors of LFT are situated on a $~$\SI{0.2}{cm} silicon wafer which is coupled to the thermal bath by an Invar clamp at the outer edges.
On the opposite side of the Si wafer, and connected to it, are $\phi = \SI{6}{mm}$ silicon lenslets. 
Because the detectors share a common thermal surface, thermal fluctuations across the wafer are effectively treated by the detectors as the thermal bath temperature $T_{0}$.

The wafer of LFT has a geometrical area of $\sim 100 \times \SI{100}{mm}$ on which 36 pixels are arranged uniformly.
Each pixel contains 4 pairs of TES detectors.
We anticipate a proton impact rate of 5 \si{cm^{-2}.s^{-1}}, or (scaled by the wafer surface area) $\approx400$ particles s$^{-1}$ into the silicon wafer. Although the spatial profile of heat dissipation in the wafer is expected to be a position-dependent effect, we expect the additive thermal fluctuation from common-mode wafer impacts to create a baseline white noise level seen by all of the detectors on one wafer.
This white noise effect is due to the strong superposition of low-level thermal fluctuations common to many detectors which are individually not resolvable with LiteBIRD's low sampling rate.
In addition to the common-mode noise described above, a second CR pulse population arises due to direct, but infrequent, impacts with the TES detectors.
However, in spite of the small size of individual TES,
direct impacts on a given TES absorber produce large-amplitude events, and are expected to occur every $\approx 160$ seconds.
We have simulated both effects in this paper.

The time-ordered data (TOD) recorded aboard the LiteBIRD spacecraft will be sampled at a rate of \SI{20}{MHz} before being processed by a series of digital filters which work as a low-pass decimation filter by on-board electronics and CPUs (respectively) to a final sampling rate of \SI{19}{Hz}.
It is important to note that the TES response is of the order of \SI{3}{ms}, thus the electrothermal response of the detectors and the thermal response of the detector wafer itself will be faster than its resolution in the final LiteBIRD TOD.
This results in the demands of LiteBIRD data post-processing to be radically different to that of Planck HFI, in which individual cosmic ray pulses were able to be distinguished.
Furthermore, the high rate of impacts into the detector wafer will create a signal dominated by the superposition of many concurrent cosmic ray impacts.
A detailed quantitative comparison between the LiteBIRD case and that of Planck HFI has been presented in prior work~\cite{tominaga2020simulation}.

Finally, LiteBIRD has a continuously-rotating Half-Wave Plate (HWP) which modulates the incoming signal and allows for the separation of $I$, $Q$, and $U$ Stokes parameters through demodulation.
In the presence of an extra intensity on the modulated signal (by, e.g., cosmic ray energy injection), this intensity becomes an additional term in $Q$ and $U$, impacting the determination of the $Q$ and $U$ Stokes parameters.
Regardless of the presence of a rotating HWP, a mismatch between the signal of two detectors in one pair increases the temperature difference of that pair, leading to leakage into $Q$ and $U$.
The magnitude of this disparity in relation to $Q$ and $U$ is the open question, and becomes especially important given the low level of the polarised signal of primordial $B$-modes.


We will describe the simulation methodology, including the physical assumptions construction of various sub-models in section~\ref{sec:simdesc}. We then discuss the results generated by the physical model, its statistical attributes, the methods we use to generate 3 years of data, and the subsequent maps and angular power spectra in section \ref{sec:postprocc}.
In section~\ref{sec:dicsandinterp} we interpret these results within the framework of the LiteBIRD space mission, as well as propose future work to improve the model and the projected outcomes of the study.

\section{Simulation methodology}
\label{sec:simdesc}
In order to address the full CR effect in LiteBIRD using simulations, it first is necessary to account for the effect of the environment of the telescope using predicted proton flux at L2 during the time of LiteBIRD's operation.
Second, we deduce the thermal response of the detector wafer with respect to its geometry as well as the attributes of the coupling of the silicon wafer to the thermal bath via the Invar holder.
These results are then used to generate simulated TOD of cosmic ray thermal fluctuations, which are finally used to generate $I$, $Q$, and $U$ maps of the cosmic ray effect.
From these maps, we derive the $C_{\ell}^{TT}$, $C_{\ell}^{EE}$, and $C_{\ell}^{BB}$ cosmic ray angular power spectra.

\subsection{Predicted flux at L2}
\label{sec:predflux}
Bearing in mind the effect of fluctuating solar magnetic fields on the proton flux~\cite{gleeson1968solar}, we have followed the methodology of ref.~\cite{lotti2018estimates}, which has been used to predict the radiative environment for the upcoming X-IFU instrument aboard the Athena telescope at L2.
As in ref.~\cite{lotti2018estimates}, we use the data from the PAMELA experiment~\cite{casolino2008launch,picozza2007pamela} which was launched in 2009 and took cosmic ray spectral measurements until February 2016.
This analysis serves the purpose of producing time-dependent flux predictions as well as overcoming underestimates of CR flux by the commonly-employed CREME96 model~\cite{davis2001evolution}.
PAMELA data is available in one month increments.

We have chosen to use Planck-era (end of 2009) spectra as the GCR flux was exceptionally high during these times, and the period was a solar minimum as LiteBIRD will be.
We take this as a worst-case scenario for which the mission should be prepared, and is the subject of our simulations.
As in the Athena case~\cite{lotti2018estimates}, we fit the PAMELA end-2009 spectra with a modified Usoskin model~\cite{usoskin2001dependence,usoskin2005heliospheric}, in which the cosmic ray differential spectrum is described using a time-dependent solar modulation parameter $\phi$.
Refs.~\cite{usoskin2001dependence, usoskin2005heliospheric} describe the cosmic ray differential intensity $J$ as
\begin{align}
\label{eq:Jeq}
J(E_{k}, \phi) &= J_\mathrm{LIS}(E_{k} + \phi) \frac{
        (E_{k})(E_{k} + E_{p} )
    }{ 
        (E_{k} + \phi)(E_{k} + \phi + E_{p})
    },
\end{align}
where $J(E_{k}, \phi)$ is at \SI{1}{AU}, $E_{p}$ is the rest mass energy of a proton (\SI{938}{MeV}), and $E_{k}$ is kinetic energy of the cosmic nuclei.
$J(T, \phi)$ is scaled by a factor of $J_\mathrm{LIS}$, representing the local interstellar spectrum (LIS) of protons, defined as
\begin{align}
\label{eq:JLIS}
    J_\mathrm{LIS} &= \frac{1.9\times 10^{4} P(E_{k})^{-2.78}}{1 + 0.4866 P(E_{k})^{-2.51}},
\end{align}
where $P$ = $\sqrt{E_{k}(E_{k} + 2E_{p})}$.

\begin{figure}[tbp]
\centering 
\includegraphics[width=0.9\textwidth]{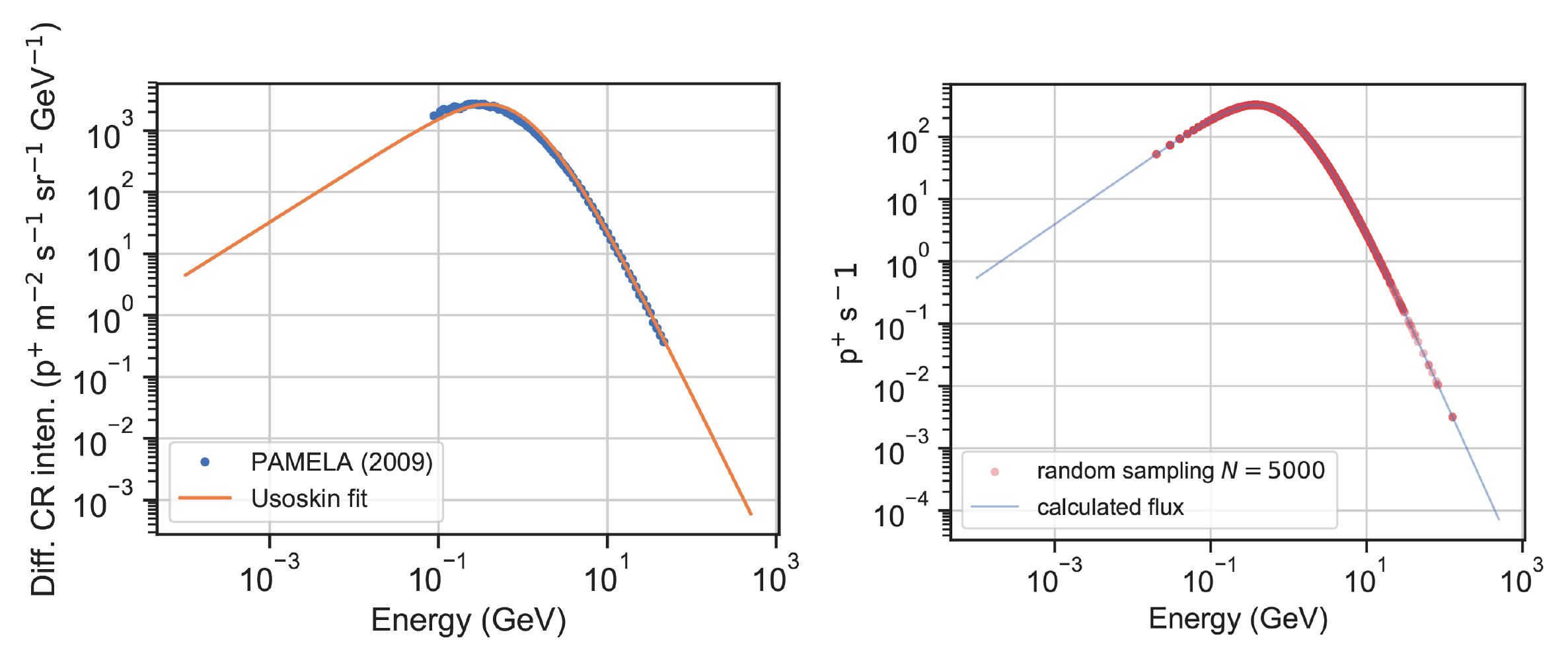}
\caption{\label{fig:PAMELAfit} \textit{Left}: Pamela end-2009 differential cosmic ray intensity (blue dots) and Usoskin model fit (orange) extended to low energy regime.
\textit{Right}: Probability density function of CR flux arriving at the detector wafer (blue line) and $N = 5000$ random sampling.}
\end{figure}

Accordingly, we fit to the PAMELA end-2009 spectra with eq.~\eqref{eq:Jeq} and eq.~\eqref{eq:JLIS} using the MPFIT $\chi^{2}$ minimisation routine~\cite{markwardt2012mpfit}, extending the lower energy boundary to \SI{1}{MeV}, as shown in figure~\ref{fig:PAMELAfit} (left).
As a first approximation, we convert the differential cosmic ray intensity to flux on the detector wafer by assuming a solid angle of $4\pi\,\si{\steradian}$, consistent with a planar detector, and multiply this by the horizontal wafer surface area $A$.
The resulting probability density function is shown in figure~\ref{fig:PAMELAfit} (right), which is later used to generate CR event tables in the TOD generation algorithm (section~\ref{sec:maketod}).

\subsection{Detector wafer thermal model}
In order to characterise the thermal behaviour of the LiteBIRD detector wafer and eventually simulate realistic TOD, we produce a model in the commercial software COMSOL \cite{multiphysics1998introduction}.
COMSOL uses finite-element modelling methods, and we use the Heat Transfer Module to assess the propagation of heat across the wafer at varying energies and energy deposition locations. Studies of this nature have been successfully performed for CR studies in other space instruments, notably Athena X-IFU, in which the impact of thermal fluctuations in the detector wafer on instrument energy resolution were assessed~\cite{peille2020quantifying,stever2020thermal,miniussi2020thermal}.

\subsection{Thermal model construction and assumptions}
\begin{figure}[tbp]
\centering 
\includegraphics[width=0.9\textwidth]{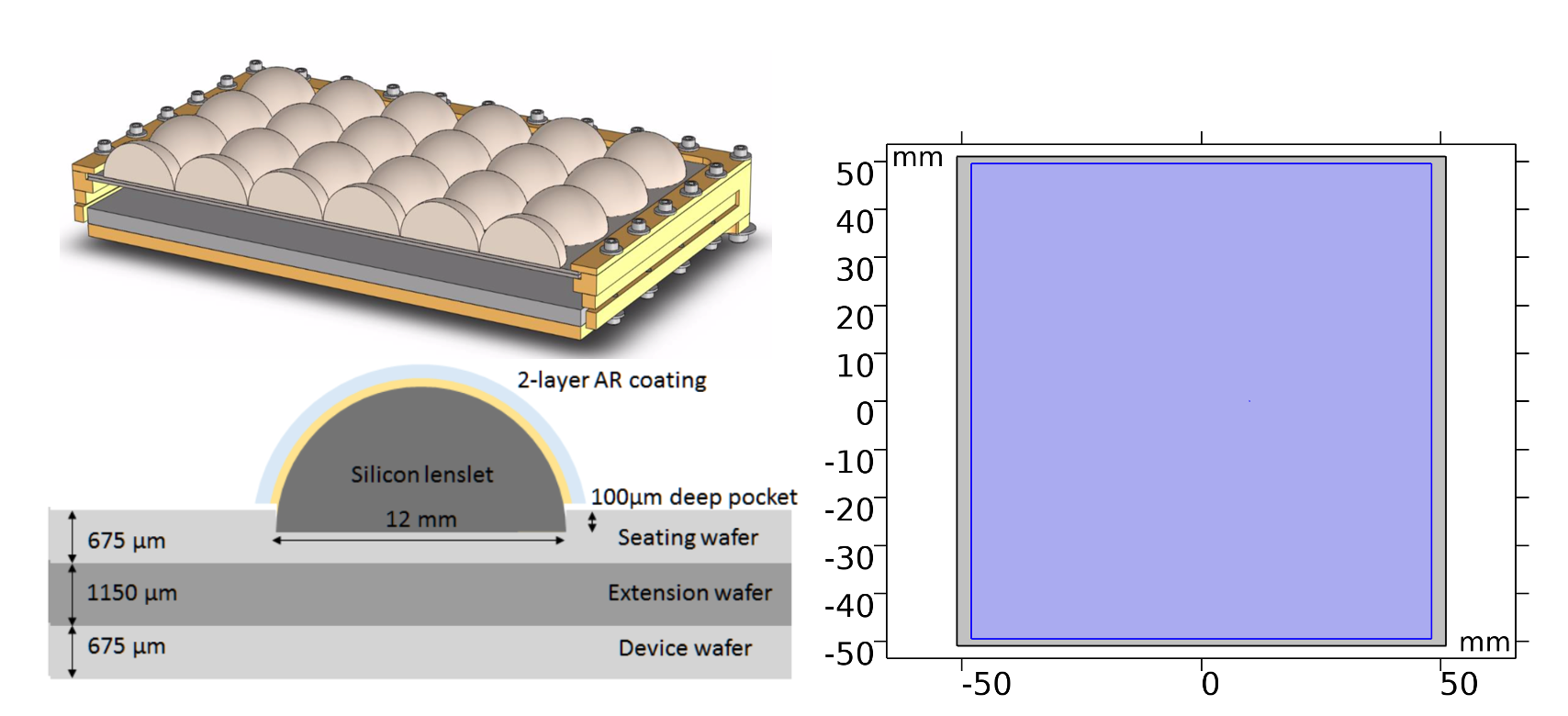}
\caption{\label{fig:stroopwafel} \textit{Left}: LiteBIRD LFT wafer design, 3-dimensional (top) and lateral (bottom). \textit{Right}: Thermal model geometrical design.}
\end{figure}

The thermal model is a two-dimensional finite-element model with a virtual $z$-axis (using the same process as described in prior work~\cite{stever2020thermal}.
However, unlike the X-IFU case, the LiteBIRD model consists only of one thermal layer due to the lack of a metallisation on the surface, i.e. the detector wafer consists only of bulk silicon.
Within the singular thermal layer, the geometry is split into two material solids, the largest for the bulk silicon and the periphery for the Invar holder.
Owing to the 2D nature of the model, we have assumed a thickness for the Si wafer based on the design specifications and added an equivalent thickness of \SI{1.7}{mm} Si from equally flattening the 36 Si lenslets.
The design of the wafer is shown in figure~\ref{fig:stroopwafel} (left), and the equivalent design of the thermal model is shown in figure~\ref{fig:stroopwafel} (right).

The Si and Invar solids have separated thermal properties, specifically thermal conductivity $k$ and heat capacity $C_{\nu}$.
In both cases $k$ goes as:
\begin{align}\label{eq:comsolk}
    k &= \frac{1}{3} \nu \lambda C_{\nu},
\end{align}
where $\nu$ is mean speed of sound in the material, $\lambda$ is the phonon mean free path, and $C_{\nu}$ is the lattice heat capacity. 

Because the Invar layer is several times thicker than the Si layer, we avoid using separate electron and phonon layers for the Invar in contrast to the treatments in refs.~\cite{peille2020quantifying,stever2020thermal,miniussi2020thermal}.
The thermal conductivity in the Invar therefore becomes the sum of the electron and phonon components, with the electron component dominating, described by a simple $T$-scaling relationship which utilises the Weidemann-Franz law and Lorenz number:
\begin{align}
\label{eq:comsolkinv}
    k_\mathrm{Inv} &= L \sigma_\mathrm{Inv} T d_\mathrm{Inv},
\end{align}
where $L$ is the Lorenz number, $d_{\mathrm{Inv}}$ is the thickness of the Invar holder, and where the electrical conductance $\sigma_{\mathrm{Inv}}$ is described as:
\begin{align}\label{eq:comsolsigma}
    \sigma_\mathrm{Inv} &= \frac{RRR_\mathrm{Inv} }{\rho_\mathrm{Inv} },
\end{align}
or the Residual Resistance Ratio ($RRR$) of the material divided by its room temperature resistivity.
The parameters for Invar have been taken from ref.~\cite{fickett1982electrical} whilst those for silicon were taken from ref.~\cite{ashcroft1976solid}.

For the heat capacity, parameters have again been taken from
ref.~\cite{ashcroft1976solid},
and scale with $T^3$ in the case of Si, and $T$ (electron) and $T^{3}$ (phonon) for Invar.
For Invar, $C_{\nu}$ is defined as the sum of the phonon and electron components:
\begin{align}
\label{eq:comsolCp}
    C_{\nu} = (C_{\nu \mathrm{phon}} T^{3}) + (C_{\nu \mathrm{elec}} T).
\end{align}

We simulate the deposited energy from a cosmic ray as an energy pulse with a rise time of 5$\times$10$^{-9}$ s and a decay time of 1$\times$10$^{-7}$ s.
This procedure is performed due to difficulties with COMSOL to process delta functions, and in the same fashion as ref.~\cite{stever2020thermal}.
This pulse is then normalised to the chosen deposited energy.
The $x$ and $y$ location of the energy deposition, as well as the deposited energy, are specified as input values. 

The model is meshed with a fine free triangular model, and it is re-meshed automatically with each position variation of the cosmic ray energy deposition.
The first stationary study step excludes the energy deposition event and calculates the ambient temperature of the wafer.
The second time-dependent solver includes the CR energy injection, and uses adaptive (uneven) time sampling with a relative tolerance of \SI{5}{ns} to resolve the thermal fluctuation at a given probe point on the wafer.
These adaptive measures allow for the resolution of the pulse profile in the time domain under a range of starting conditions, allowing for the assessment of the Noise Equivalent Power (NEP) in the noise spectra.

In the baseline design for LiteBIRD, coupling of the wafer to the thermal bath is via an Invar holder into which the detector wafers are clamped.
In our thermal simulation, we therefore assume that Kapitza-like thermal boundary resistance is the dominant form of thermal coupling, in the absence of additional wirebonds.
The definition of this thermal transfer is simply:
\begin{align}
\label{eq:comsolkapitza}
    Q_{0} = 2 K (T_{0}^{4} - T_{\mathrm{Inv}}^{4}),
\end{align}
where $T_{0}$ is the LiteBIRD bath temperature of \SI{100}{mK} and $T_{\mathrm{Inv}}$ is the temperature of the Invar in a given mesh pixel.
$K$ is the Kapiza coupling constant between Si and Invar, which can be variable depending on the clamping force applied.
This value has not yet been measured in a fabricated LiteBIRD wafer, so a heritage value measured for a copper-silicon boundary for the wafer design of the SPICA mission~\cite{stever2020thermal,peille2020quantifying} has been used.
$K$ determines the speed with which thermal energy is transferred across the wafer surface, which is of significant importance to the outcome of the simulations.
Its laboratory measurement is consequently given priority in future LiteBIRD experimental work.

\subsection{Thermal wafer model output}

\begin{figure}[tbp]
\centering 
\includegraphics[width=0.7\textwidth]{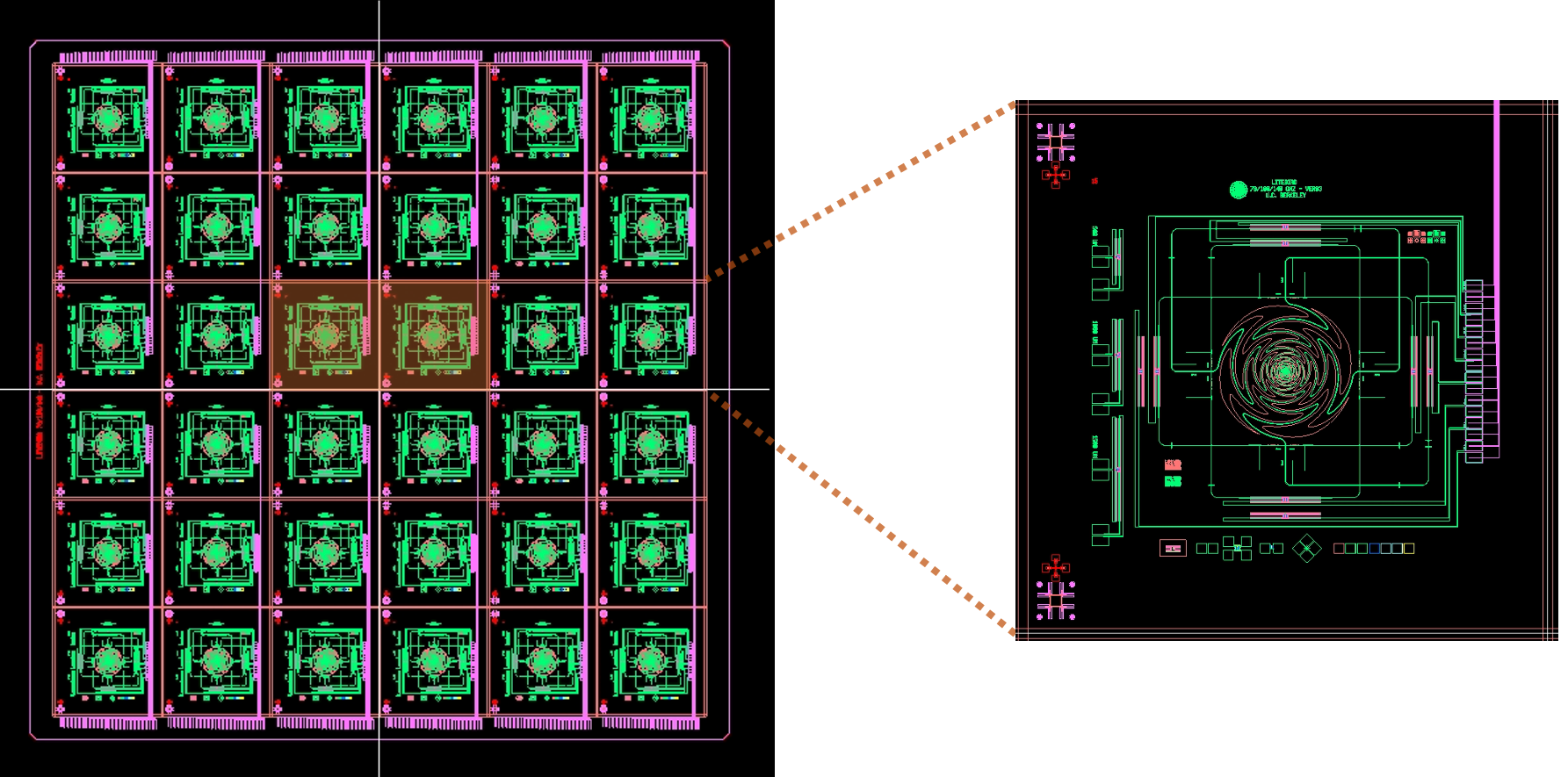}
\caption{\label{fig:detlayout} Design of LiteBIRD wafer and pixels, with 2 simulated pixels indicated (orange) \cite{suzuki2013multichroic}.}
\end{figure}

The output of the thermal model is $T (t)$ timestreams calculated at the surface (using temperature probes) of the Si substrate for a given amount of deposited energy, as well as its location.
For the end-to-end analysis described in this work, we have produced a library of $T (t, E, x, y)$ in 16 locations in which we place the 16 simulated TES.
This area comprises 2 central pixels with 8 detectors each, in the orientation shown in figure~\ref{fig:detlayout}.

\begin{figure}[tbp]
\centering 
\includegraphics[width=0.7\textwidth]{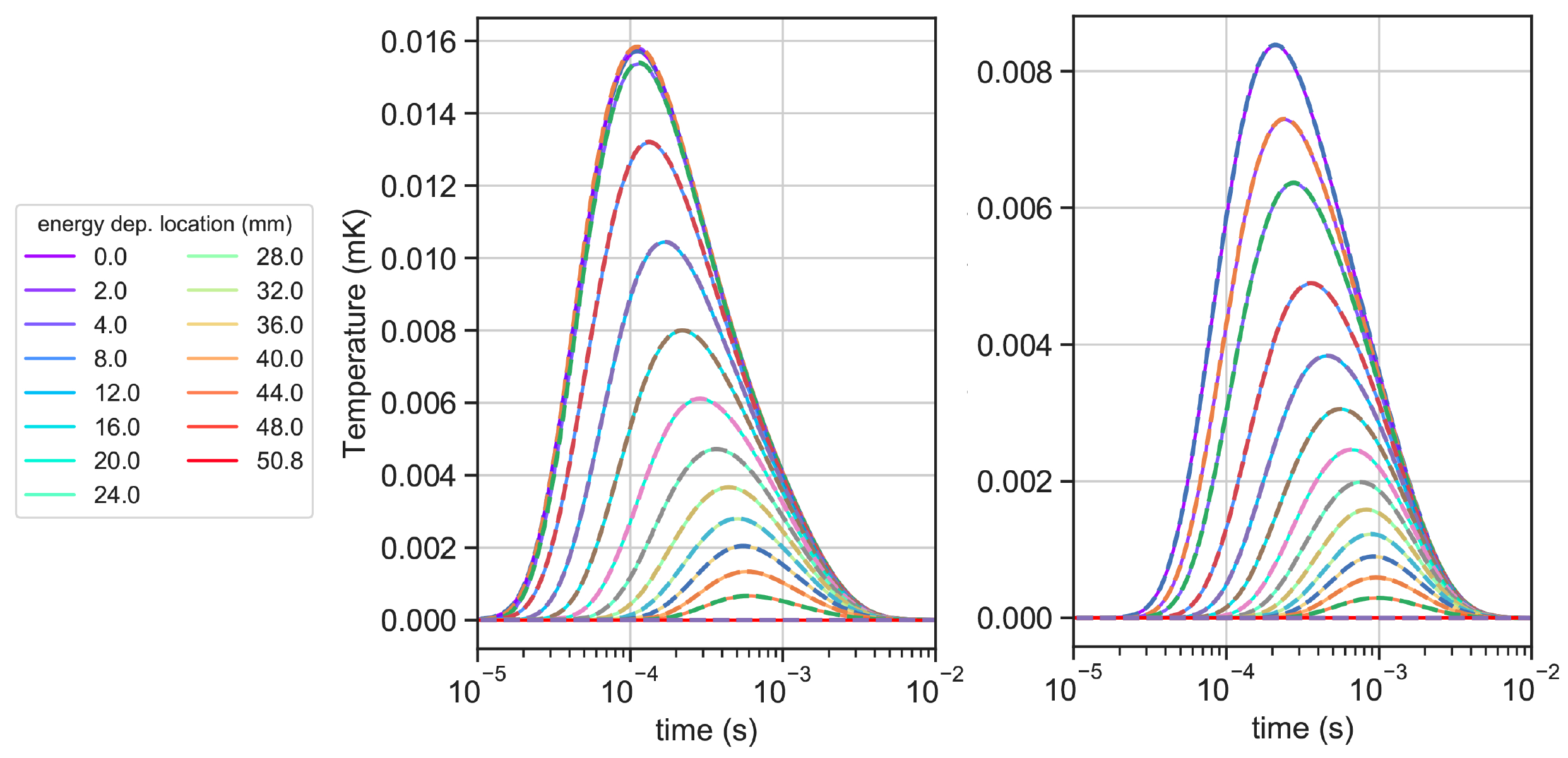}
\caption{\label{fig:comsolTscale} $T (t, E, x, y)$ wafer surface temperature timestreams in the location of TES 9 (left) and TES 2 (right).
The solid lines indicate the thermal response to a deposited energy of \SI{0.5}{MeV}, and the dashed lines indicate timestreams of \SI{5}{MeV} energy depositions in the same locations, scaled to \SI{0.5}{MeV}.}
\end{figure}

Analysis of these wafer response timestreams has determined that they are easily scalable for the employment of a TOD-production code in terms of the wafer thermal response at the energy deposition range of our interest.
To facilitate this scaling, we calculate the relationship between deposited energy and pulse amplitude, and use the interpolated energy-amplitude relationships at the distance of each deposition to the thermal probe to predict the pulse height for any deposited energy, using the same procedure described in refs.~\cite{stever2020thermal,peille2020quantifying}.
An example of this is shown in figure~\ref{fig:comsolTscale} (solid lines) in which \SI{0.5}{MeV} is deposited at various distances from the wafer centre and temperature is measured at the locations of TES 9 (left) and TES 2 (right).
The dashed lines in figure~\ref{fig:comsolTscale} correspond to pulses from \SI{5}{MeV} depositions, downscaled to \SI{0.5}{MeV} pulse heights using the scaling procedure.

Three $T (t, E, x, y)$ pulse sets at $0.5$, $50$, and \SI{5}{MeV} at each TES location compose the master pulse library from which we generate TOD using the algorithms described in the next section.

\subsection{TOD generation algorithm}
\label{sec:maketod}
Generation of detector TOD is composed of two elements:
\textit{i.} the generation of an event list,
and \textit{ii.} generation of TOD from the event list.
Whilst it is achievable to randomly generate events at the same time as generating TOD, we keep these aspects relegated to separate algorithms to save computation time and to separately preserve a single event table for later analysis.

For the purposes of the work outlined in this manuscript, we generate 90 minutes of event tables and 90 minutes of TOD.
Assuming $\approx400$ events \si{s^{-1}}, following Poisson statistics, we generate random arrival times from an exponential distribution of the time difference between two sequential events \cite{VerabschiedetdurchdieUniversitaetsleitung2011}.
An arrival energy is drawn from the probability distribution function of incoming CR energies described in section~\ref{sec:predflux}.
For our simple first-order estimate, we assume a low energy cutoff of \SI{50}{MeV} resulting from the aluminium payload of the satellite, although the real case will involve variable thickness of mostly Al spacecraft hull and focal plane components.
From the arrival energy, we generate a random striking angle $\theta$, and from this we calculate the thickness of Si traversed by the CR and calculate the deposited energy as a function of the $dE/dx$ from stopping power tables.
Random $x$ and $y$ locations on the wafer surface are then populated.
The event tables are recorded as 90 minute lists of arrival energy $E_{\mathrm{draw}}$, deposited energy $E_{\mathrm{dep}}$, striking angle $\theta$, arrival times $t_{a}$, and $x$ and $y$ locations.

The separate algorithm for generating TOD loads the event tables, loops over the number of TES simulated, and loops over each event in the event list.
We calculate the distance of the deposited energy from the current TES, and choose a suitable pulse from the $T (t, \delta E, \delta x, \delta y)$ library depending on the necessary location and energy.
This library pulse is normalised and scaled to an appropriate amplitude and added to the TOD array at a time corresponding to the arrival time $t_{a}$.
At the end of each event loop, we calculate whether the incident cosmic ray impacts the geometrical area of the TES absorber itself.
If this condition is satisfied, we calculate the energy deposited into the palladium absorber ($E_{\mathrm{dep-TES}}$) with an assumed thickness of $\approx$1 $\mu$m as a function of $E_\mathrm{draw}$, $\theta$, and $dE/dx$ for Pd.
The process is repeated for the much smaller aluminium thermistor if it is within the impact region, assuming the same thickness.
In this case $dE/dx$ is calculated from the stopping power table for silver due to palladium being unavailable on PSTAR \cite{berger1992estar} and the atomic commonalities between Pd and Ag.
The final $E_{\mathrm{dep-TES}}$ is converted to power $P_{\mathrm{CR}}$ by multiplying it by the sampling rate (nominally \SI{20}{MHz} in LiteBIRD but reduced to \SI{10}{kHz} in our simulation to save computation time).
$P_{\mathrm{CR}}$ is added to the assumed optical power $P_{\mathrm{opt}}$ at the time index of the impact, thus treating CR energy injected directly into the TES as a delta function which is resolved through the electro-thermal TES response function along with the thermal fluctuation.

After looping over each event and each TES, the resultant arrays $T_{\mathrm{wafer}} (t)$ and $P_{\mathrm{TES}} (t)$ (wafer surface temperature and power incident on the TES, respectively) are processed by a function for determining the TES response.

\subsection{TES response calculation}
\label{sec:tomma}

We model the TES response according to ref.\cite{Irwin2005} through the following differential equations, where we assume a DC-biased bolometer:
\begin{align}
    C\frac{dT}{dt}=-P_b+P_{\mathrm{opt}}+P_{\mathrm{el}},
    \label{eq:thermal_diff}
\end{align}
\begin{align}
    L\frac{dI}{dt}=V-R_{\mathrm{TES}}I-R_{s}I , 
    \label{eq:current_diff}
\end{align}
where $T$ and $I$ are respectively the temperature of the bolometer island and the current flowing through the thermistor.
The other quantities appearing in eq.~\eqref{eq:thermal_diff} and eq.~\eqref{eq:current_diff} are the heat capacity of the bolometer $C$, the power $P_b$ flowing along the weak link between the bolometer island and the thermal reservoir at temperature $T_b$ (the nominal value would be \SI{0.1}{K}, however this is affected by the impact of cosmic rays on the focal plane), the optical power loading the absorber $P_{\mathrm{opt}}$ and the electrical bias power $P_{\mathrm{el}}$ which is equal to $V^2/R_{\mathrm{TES}}$ for a voltage-biased bolometer of resistance $R_{\mathrm{TES}}$.
Lastly, $L$ and $R_s$ are the SQUID-input inductor and the shunt resistor ($R_s\ll R_{\mathrm{TES}}$) that maintains the voltage bias condition.
We have adopted representative values for $L=\SI{60}{\mu H}$ and $R_s = \SI{0.02}{\ohm}$ to match the parameter expected for readout that will be employed by LiteBIRD.

In order to define the other relevant parameters we assumed a representative optical loading power of \SI{0.5}{pW} (although depending on the frequency channel we expect the value to be in the range $\SI{0.2}{pW} \lesssim P_{opt}\lesssim \SI{0.6}{pW}$, details of which can be found in
ref.~\cite{westbrook2020detector},
and an intrinsic thermal time constant $\tau_0=C/G\sim\SI{33}{ms}$, where $G$ is the thermal conductance of the weak link between the TES island and the thermal reservoir. 

To model $P_b$ we again assume the classical formula~\cite{Irwin2005}:
\begin{equation}
    P_b = \frac{G}{nT^{n-1}}(T^{n}-T_b^{n}),
    \label{eq:Pb}
\end{equation}
where $n = \beta+1$ and $\beta$ depends on the primary thermal carriers (in this work we assume $\beta=3$ for phonons).
By design the LiteBIRD detectors will target a saturation power $\sim 2.5\times P_{opt}$.
From these parameters we can derive the values of $G$ and $C$ for the expected saturation power.
In the case of $P_{opt}=\SI{0.5}{pW}$, we find $G=\SI{33.1}{pW/K}$ and $C=\SI{1}{pJ/K}$.

The target normal resistance $R_n$ for LiteBIRD TES detectors is $R_n \sim \SI{1}{\ohm}$, therefore we created an analytical model using an $\arctan$ approximation to mimic the behaviour of the superconductive transition ($R_{\mathrm{TES}}(T)$).
The analytical model is created assuming that the width of the transition is \SI{10}{mK} (defined as the temperature window where the resistance varies in the range $\SI{0.25}{\ohm} < R < \SI{0.75}{\ohm}$).
With this assumption we obtain a value of $\alpha\sim100$ (the logarithmic derivative of $R$ with respect to $T$) at the center of the transition ($R\sim \SI{0.5}{\ohm}$ at $T=\SI{171}{mK}$).
More details about the assumptions can be found in
ref.~\cite{ghigna2020a}.

In order to simulate the response to cosmic rays, the bath temperature fluctuations produced with the TOD generation (section~\ref{sec:maketod}) are passed to a 
\texttt{python}
routine\footnote{\url{https://github.com/tomma90/tessimdc}} which solves the coupled differential equations using the Runge-Kutta 4th-order method (details can be found in ref.~\cite{ghigna2020a}), which yields detector response arrays in current.

Lastly, we assume that we will be able to operate the TES detectors in high loop-gain regime, $\mathcal{L} \sim 10$.
Therefore, when we simulate the detector response we tune the electrical bias to fix the operation point at this level.
In this regime the current responsivity can be easily computed as $S_{I}=-1/V$. We use this value to convert the current output to equivalent power at the input of the bolometer.

\subsection{TOD filtering, decimation, and power conversion}
The resulting array of TES current ($I (t)$) is decimated in two stages in accordance with the current LiteBIRD design specifications.
The first decimation stage is by Cascaded Integrator–Comb (CIC) filter with a decimation ratio $R = 1/64$.
The decimated current arrays are then converted to power by dividing them by the current responsivity calculated using the algorithm described in section~\ref{sec:tomma}.

\section{TOD analysis, postprocessing, and mapmaking}
\label{sec:postprocc}
In this section we describe the final processing stages of the TOD generated by the physical model, as well as its statistical attributes in section \ref{sec:analyseTODphys}. 
We then present a template-injection method for extending the physical model TOD from 90 minutes up to the mission duration of 3 years in section \ref{sec:analyseTODtemp}. 
Finally, we describe the method of projecting this TOD to sky maps and subsequent power spectra in section \ref{sec:howtomakemap}.

\subsection{TOD from physical model}
\label{sec:analyseTODphys}

\begin{figure}[tbp]
\centering 
\includegraphics[width=0.7\textwidth]{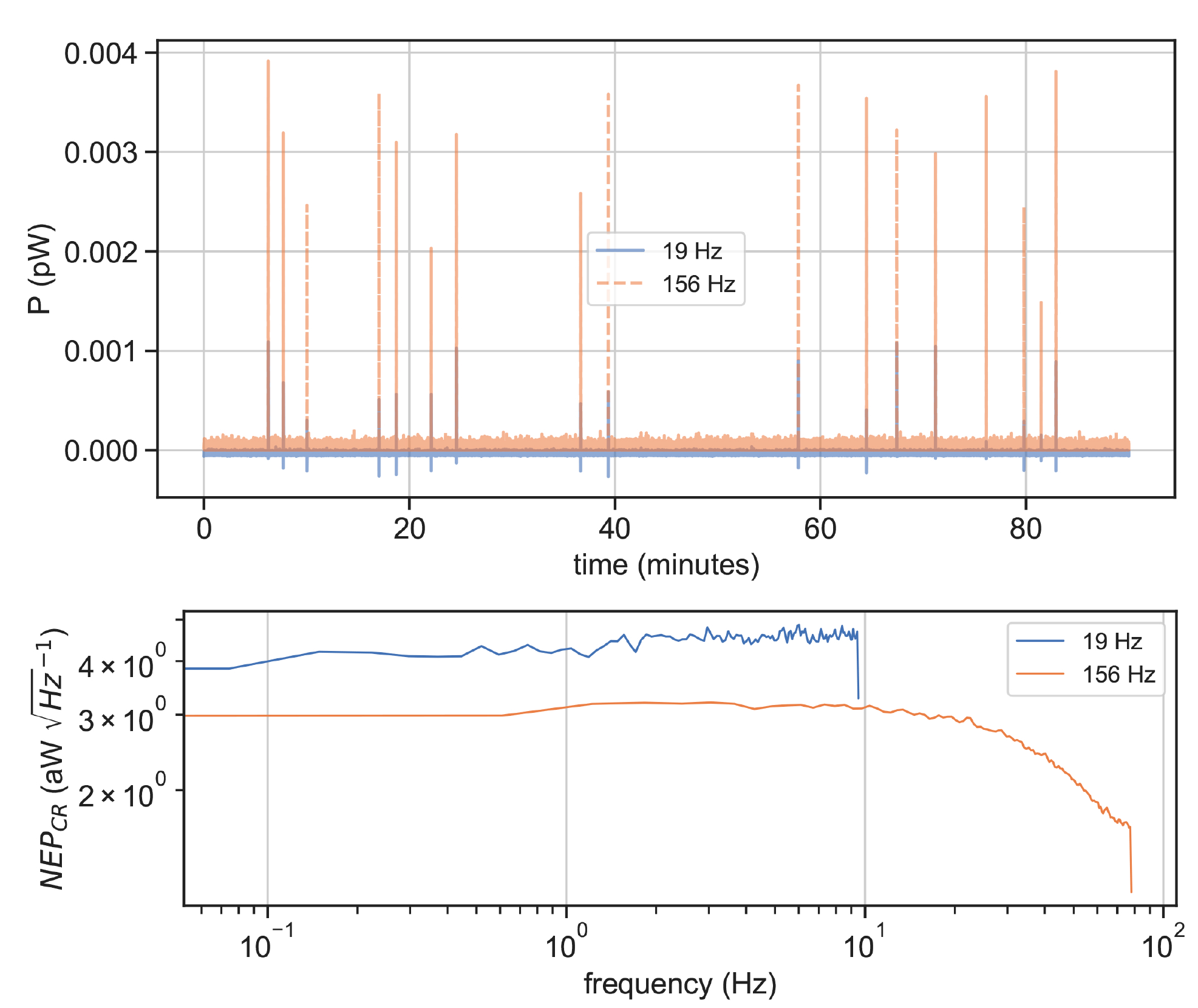}
\caption{\label{fig:CRNEP} \textit{Top:} An example of \SI{156}{Hz} and \SI{19}{Hz} TOD, with low-level thermal noise from wafer hits, and the less-common direct hits.
\textit{Bottom}: Noise power spectra of the simulated TOD from the physical model.}
\end{figure}

As mentioned in the previous section, the generated TOD contains two components:
the low-level superimposed thermal fluctuations from the large number of CR hits in the detector wafer,
and the much less common direct hits into the TES.
The direct hits occur $\approx120$ times per day, and are characterised by their high amplitudes and short time constants.
Downsampling the \SI{156}{Hz} data by FIR filter reduces the amplitude of direct hits, but introduces a negative component (ringing) due to the filter response.
From this TOD we derive a Noise Equivalent Power (NEP) of approximately \SI{10}{aW. \sqrt{Hz}^{-1}} and a relatively flat noise power spectrum, as shown in figure~\ref{fig:CRNEP}.

\begin{figure}[tbp]
\centering 
\includegraphics[width=0.65\textwidth]{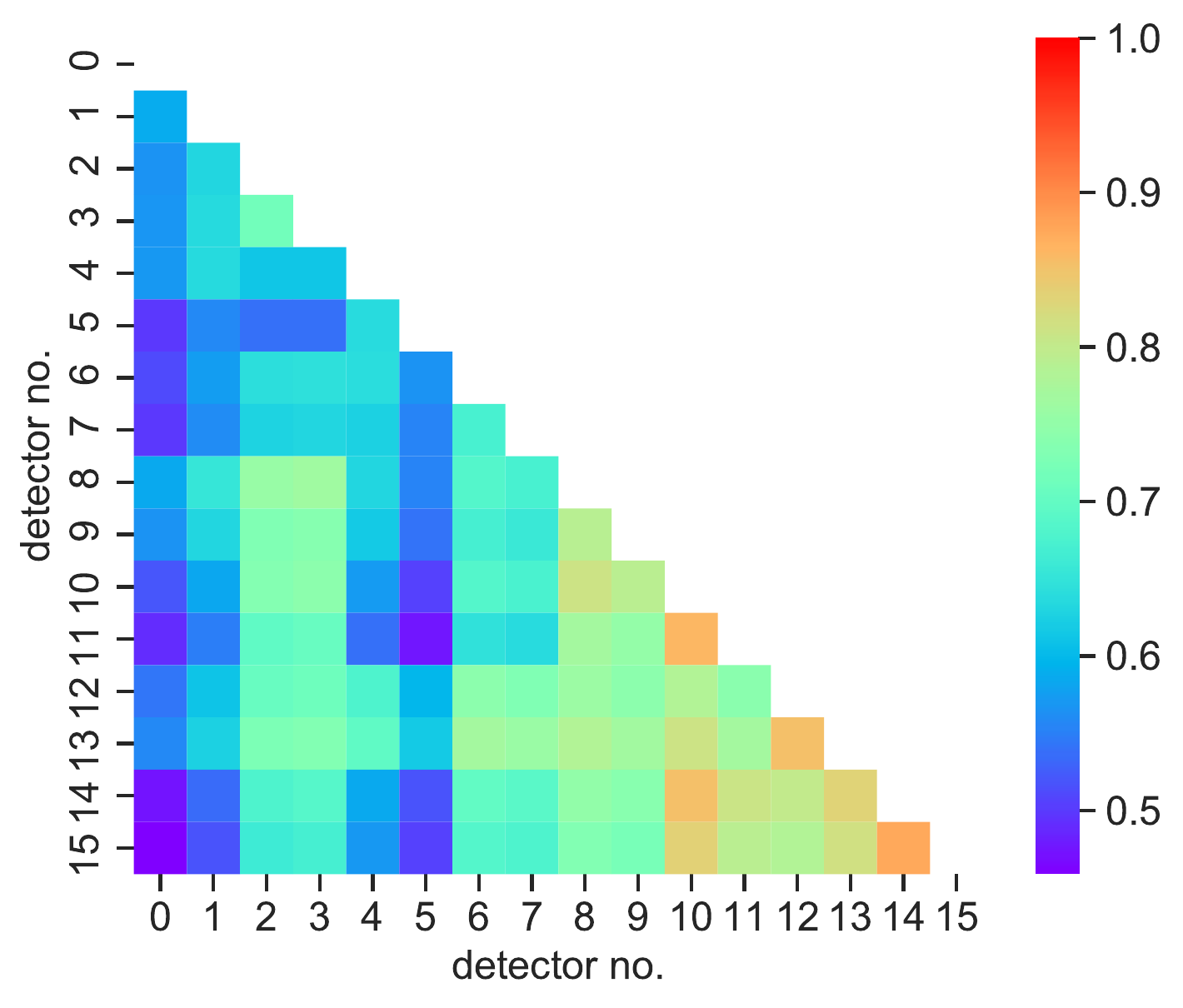}
\caption{\label{fig:TODcorro} Correlation matrix of coupling between the 16 detectors as generated by the physical model. The correlation coefficient is a value between 0 and 1.}
\end{figure}

Due to the strong common-mode thermal coupling between the 16 detectors, we produce a correlation matrix to assess the overall level of common noise between the detectors, which we show in figure \ref{fig:TODcorro}.
We observe correlations largely between 45\% and 90\% between detectors, with a mean correlation coefficient of 68.2\% and a minimum of 46.0\%.
Due to this observation, it is necessary to consider the effect of common-mode noise in further analysis.

\subsection{TOD from template injection}
\label{sec:analyseTODtemp}
As LiteBIRD uses a continuously-rotating achromatic Half Wave Plate (HWP) to modulate the polarisation of incoming radiation, there is no need for pair differentiation in the data analysis.
In this section we analyse the template-generated TOD for a single detector pair over 4 hours both from the perspective of pair summation and pair differencing, post-decimation (at \SI{19}{Hz}).

In ref.~\cite{tominaga2020simulation}, the data were shuffled to produce single-year length TOD, but this produced an artificial correlation in time, and hence was found inadequate. In light of this, they produced single-year TOD from a white noise spectrum with the same amplitude with the CR noise.
This removes the common-mode feature of the CR noise.

In this paper, we adopt an improved method consisting of template injection methods.
A statistical analysis of the 90 minute TOD is made at \SI{156}{Hz} (before the final decimation by FIR), and the derived attributes of this TOD is used for producing longer segments of data.
The low-level thermal noise produced by $\approx400$ hits per second in the wafer was found, through analysis of the TOD generated by the physical model, to be a simple white noise with strong correlation between detectors, and is injected as such into the TOD for each detector whilst preserving the common mode element derived from the correlation matrix for all 16 TESs.
For the direct CR impacts into the detector themselves, we analysed the generated 90 minutes of TOD with a peak-finder and fit an exponential profile to the individual pulses as:
\begin{align}
\gamma(t) = C_{1} + C_{2} e^{-t/\tau}, 
\label{eq:fit_glitch}
\end{align}
where $C_1$ and $C_2$ are both measured in units of power and representing respectively the low-level noise plateau and the amplitude of the direct hit; $\tau$ is the time constant aimed at characterizing the post-glitch exponential drop. 

Eq.\ref{eq:fit_glitch} is fitted on the samples encoded during the 120 ms after the identified glitch in the simulated TOD. In fact, we find that this timescale reduces extra correlations between the fitted parameters. Moreover, we flag the direct hits happening subsequently within the 120 ms, as this results in a poorly constrained fit.  

For each 90 minutes of TOD from the physical model, we can derive the distributions of the best-fit parameters $\mathcal{P}(C_1), \mathcal{P}(C_2),$ and $ \mathcal{P}(\tau)$ that can be inverse resampled to randomly inject direct hits into longer TOD simulations (e.g. 3 years) in such a way that the resampled distributions are statistically consistent with the ones observed in the 90 minute physical model data. 

\begin{figure}[tbp]
\centering 
\includegraphics[width=0.7\textwidth]{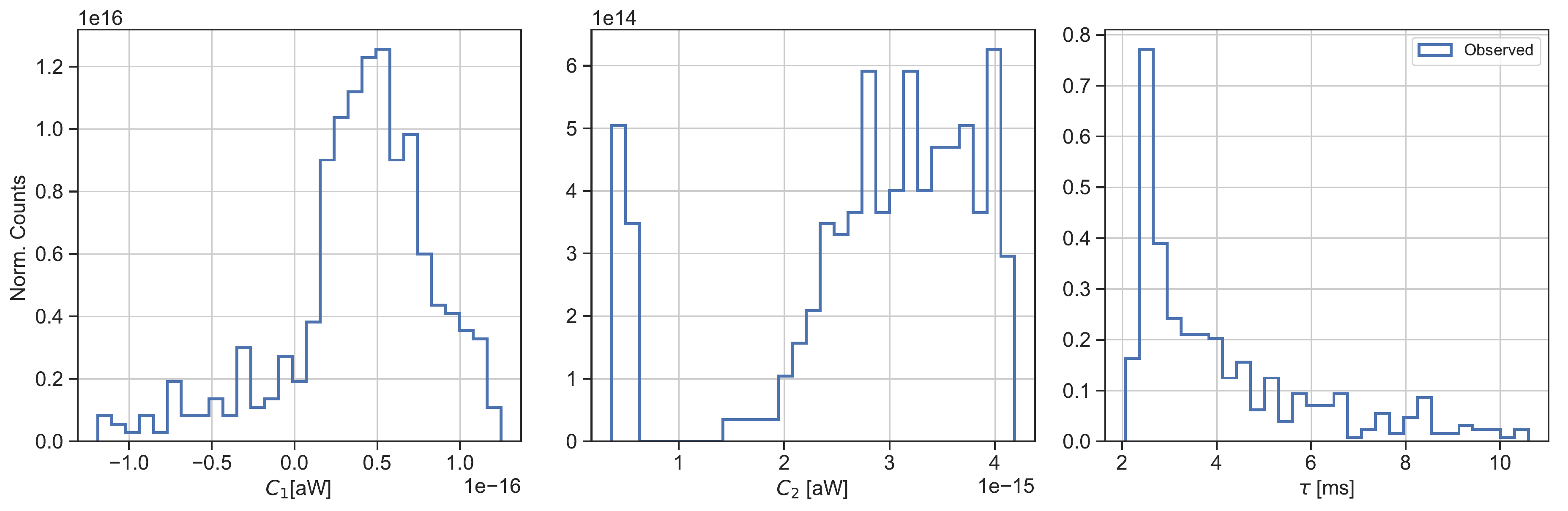}
\caption{\label{fig:fit_params} Distributions of the fitted parameters $\mathcal{P}(C_1), \mathcal{P}(C_2), and \mathcal{P}(\tau)$ used for generating CR TOD injection templates.}
\end{figure}

The CR injection algorithm implemented in TOAST is outlined for each detector as follows: 

\begin{itemize}
    \item Estimate the expected number of glitches $N_{\textrm{exp}}$ given the expected hit rate $R$=\SI{0.0014}{Hz} and the observation length (assumed hereafter $t$=\SI{24}{h}). We expect $N_{\textrm{exp}}$ $\approx$ 120 hits per day and we draw an integer $N_{\textrm{hits}}$  with a Poisson distribution peaked around $N_{\textrm{exp}}$. 
    \item Draw from a continuous distribution $N_{\textrm{hits}}$ time events within one observation timescale and inject the direct impacts  in the corresponding time samples by randomly resampling the distribution of best-fit parameters $(C_1, C_2, \tau)$. 
    \item Downsample the direct hit signal from 156 Hz down to the nominal sampling rate frequency.
    \item Coadd the direct hit timeline to the low-wafer thermal noise  one.
\end{itemize}

The TOD generated by template sampling is consistent with the TOD directly from the physical model, with NEPs also of the order of \SI{10}{aW.\sqrt{Hz}^{-1}} and with the same flat features.
Inclusion of the common-mode noise has the expected effect of producing orthogonal detector pairs with a strong thermal coupling, but with an overall signal magnitude of the same order as those generated with simple white noise.

In order to coadd the CR signal to the astrophysical signal and instrumental noise, we convert the CR TOD in Watts to CMB temperature units using the following conversion factor:
\begin{align}
\frac{dP}{dT_{\mathrm{CMB}}} = \frac{1}{k_{B}} \int d \nu \eta(\nu) \left( \frac{h \nu}{T} \right)^{2} \frac{e^{h \nu / k_{B}T}}{(e^{h \nu / k_{B}T}-1)^{2}},
\end{align}
where $B_\nu$ is the Planck blackbody function at $T$ = \SI{2.725}{K} as a function of the band frequency $\nu$, and $\eta(\nu)$ is the optical efficiency, which is defined in the LiteBIRD instrument model on a per-band basis.

\subsection{Propagation of CRs to sky maps and power spectra}
\label{sec:howtomakemap}
We perform simulations with TOAST in order to reproduce the LiteBIRD satellite nominal scanning strategy: i.e. precession and spin periods of respectively 192.35 and 20 minutes per rotation and precession axis and boresight angles respectively of 45 and 50 degrees. We set the length of the survey to 3 years and we included no astrophysical or cosmological signal to better single out the effects from CR hits. 

\begin{figure}[tbp]
\centering 
\includegraphics[width=0.7\textwidth]{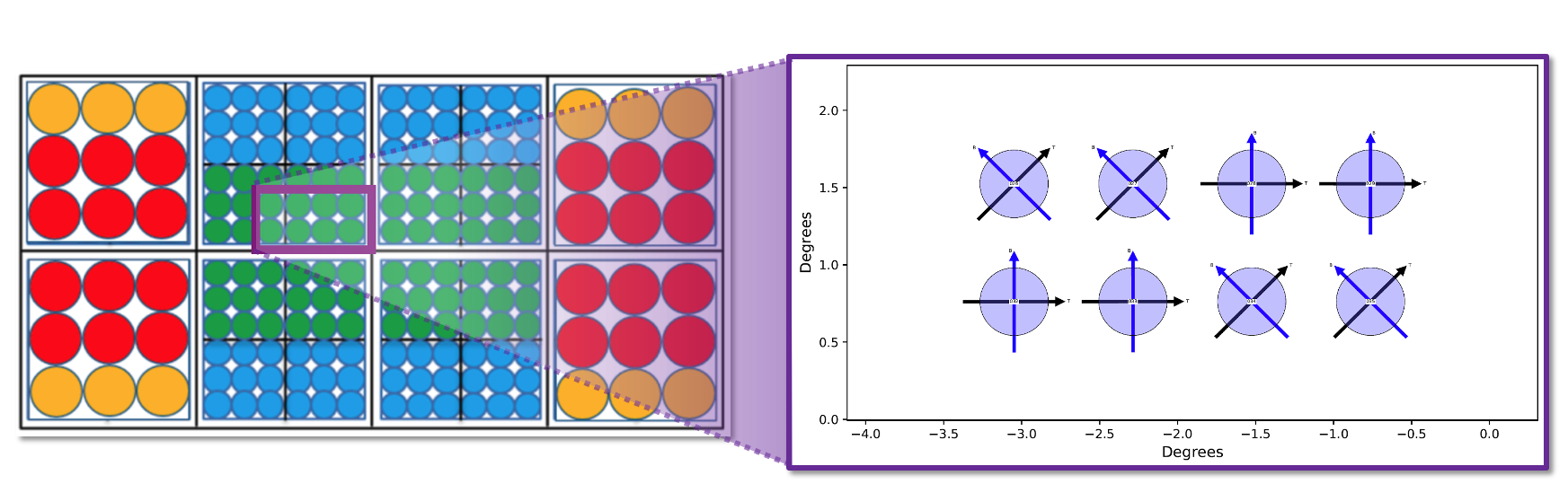}
\caption{\label{fig:detectorlayout} The layout of the simulated detectors on one wafer of the Low Frequency Telescope.}
\end{figure}

We considered 4 different configurations for the detector layout assumed on the Low Frequency Telescope (LFT) containing 1, 8, 16, and 32  detector pairs as depicted in figure~\ref{fig:detectorlayout}.
Analysis of other wavebands and other layouts, such as those on the Medium and High Frequency Telescope (MHFT) is the subject of future work.

We simulate our dataset accounting for Half-Wave Plate modulation spinning at \SI{46}{Hz} rate. 

The TOD for the entire survey is produced and projected into a HEALPix \citep{Zonca2019, 2005ApJ...622..759G} full-sky map at \texttt{nside=256} using the \texttt{libmadam} map-making library \citep{Keih_nen_2010,Keih_nen_2005}. The entire simulation is performed  with 210 KNL nodes (3360 processing elements) of the National Energy Research Scientific Computing Center (NERSC) supercomputing facilities and requires about 625 cpu hours to finalize the simulation run.

\subsection{Derived maps and power spectra}
\label{sec:results}

As a basis for comparison, we have simulated the cosmic ray signal, as outlined in section \ref{sec:analyseTODtemp}, for several hardware setups (e.g. with and without the HWP rotation) and accounting (or not) for the common-mode thermal noise.

\begin{figure}[tbp]
\centering 
\includegraphics[width=0.9\textwidth]{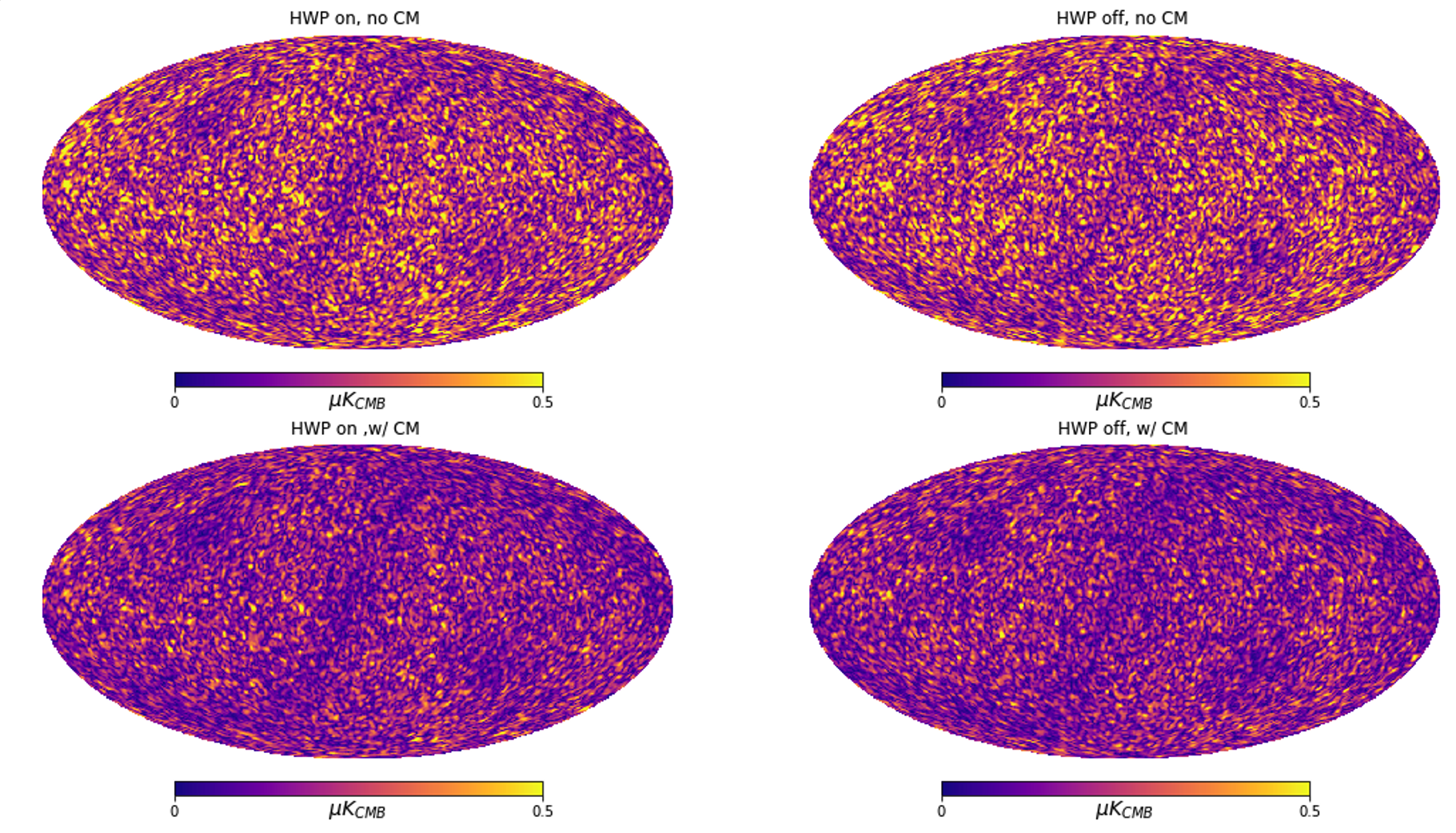}
\caption{\label{fig:skymap} \textit{Left:} Single detector sky maps of CR noise with the HWP rotation on without common-mode noise (top) and with common-mode noise (bottom). \textit{Right}: The same as above, with no HWP.}
\end{figure}

\begin{figure}[tbp]
\centering 
\includegraphics[width=0.9\textwidth]{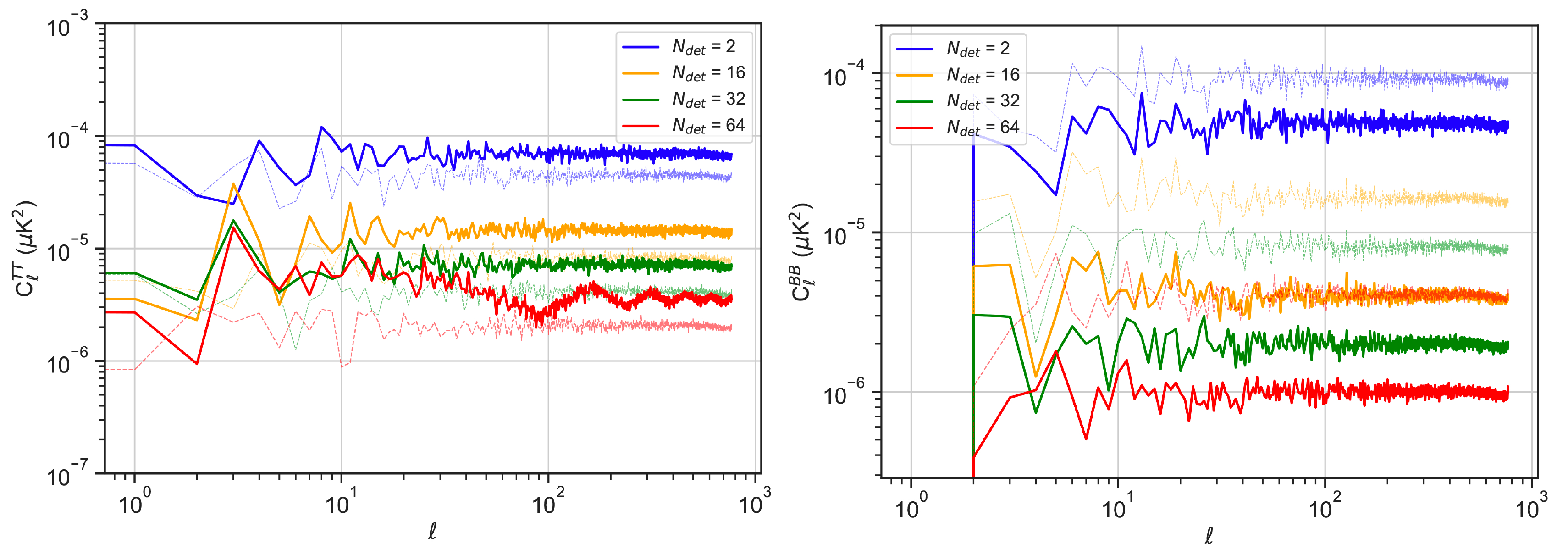}
\caption{\label{fig:CLBB64} \textit{Left:} The $C_{\ell}^{TT}$ and  $C_{\ell}^{BB}$ cosmic ray power spectra with (solid lines) and without (dashed, light colour lines) common-mode noise for 2, 16, 32, and 64 detectors on the LFT 119 GHz band.}
\end{figure}

The generated cosmic ray noise maps are shown in figure \ref{fig:skymap}. From these, we have estimated the power spectra using the HEALPix function \texttt{anafast}, from which we derived the $C_{\ell}^{TT}$ and $C_{\ell}^{BB}$ power spectra.
We find that both with and without the common-mode noise component, in the case of two detectors, the $C_{\ell}^{BB}$ CR power spectrum for two detectors is flat and with a magnitude $\approx$10$^{-4}$ $\mu$K$_{\mathrm{CMB}}^{2}$, as shown in figure \ref{fig:CLBB64}, with the former amplitude being lower than the latter.  

We repeat the above treatment for 32 detectors (16 pairs), with 8 pairs each on one wafer, as well as for 64 detectors (32 pairs) on four wafers.
The increase from 2 to 64 detectors results in a lower level of $C_{\ell}^{BB}$ power, decreasing from $\approx$10$^{-4}$ to $\approx$10$^{-5}$ $\mu$K$_{\mathrm{CMB}}^{2}$ (a factor of 10) without common-mode noise injection and $\approx$10$^{-6}$ $\mu$K$_{\mathrm{CMB}}^{2}$ (a factor of 100) with common-mode noise.\\

\section{Discussion}
\label{sec:dicsandinterp}
In this section, we discuss the results obtained in section~\ref{sec:results}, as well as some presently-achieved sensitivity calculations, implications for instrument success, and plan for future work.

\subsection{Interpretation of CR power spectra}
First, we note that the power spectra of the generated TOD, both from the output of the physical model as well as the template-injected TOD, report NEPs of the order of \SI{10}{aW} $\sqrt{\mathrm{Hz}}^{-1}$.
This NEP is also of the same order as the LiteBIRD external noise budget for a single detector in the \SI{119}{GHz} band ($\approx$\SI{11.2}{aW \sqrt{\mathrm{Hz}}^{-1}}).
It is therefore a considerable design goal to reduce this NEP using hardware mitigation as well as postprocessing of the scientific data to mitigate the CR systematic effect in LiteBIRD. 

An important consideration with respect to both the $C_{\ell}^{TT}$ and $C_{\ell}^{BB}$ power spectra is that the intensity and polarisation both scale down with the number of detectors, but that $C_{\ell}^{TT}$ with common-mode noise is higher than without it, and that this is inverted with $C_{\ell}^{BB}$.
This implies that the $B$ mode sensitivity forecasting, for a given CR intensity, is more optimistic as the thermal white noise becomes more correlated. 
This phenomenon is driven by the projection of HWP-modulated TOD into maps, where the HWP modulation acts like an effective pair difference between two detectors in a single pair.

\subsection{Thermal coupling within detector pairs}
We surmise that the above effects are due to effective pair-differencing arising from the mapmaking procedure - the individual detector TOD of a pair are not explicitly pair-differenced in the maps, but two orthogonal detectors are assigned to opposite-sign weights in the \texttt{libmadam} mapmaking framework due to the fact that the two bolometers within a detector pair are assumed to be orthogonally-oriented and that the TOD of each is weighted with same noise weight. 
As a result, the case with strong thermal coupling between a single pair of detectors will produce polarisation maps with an overall averaging effect, and thus a lower intensity than the case  with simple statistically-independent CR noise.

\begin{figure}[tbp]
\centering 
\includegraphics[width=0.8\textwidth]{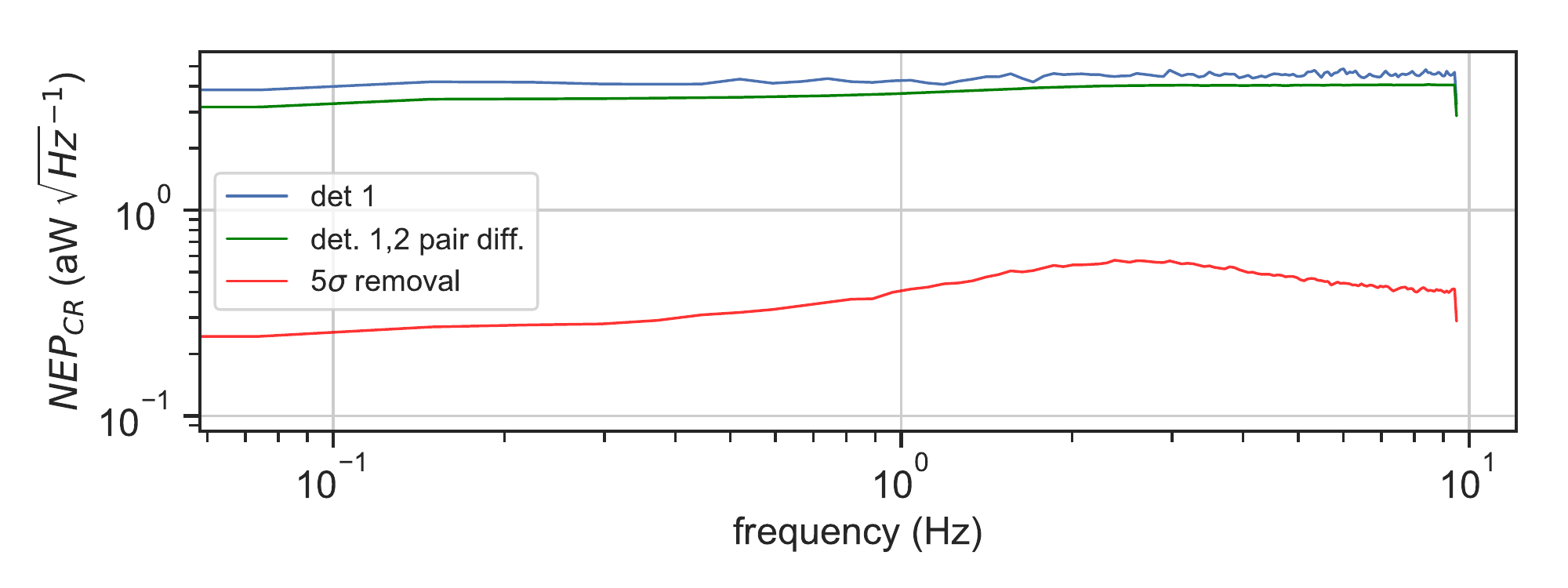}
\caption{\label{fig:TOD_psd_diff} Power spectra of single detector TOD (blue), pair-differenced TOD of detectors 1 and 2 (green), and pair-differenced TOD with a 5 $\sigma$ deglitching filter (red).}
\end{figure}

Furthermore, this effective pair difference leads to the importance of simple deglitching in order to reduce the CR signal intensity; although direct TES impacts are infrequent, they have a large amplitude which is restricted to a single detector. As an example, we employ a simple $5 \sigma$ deglitching filter on 90 minutes of pair-differenced TOD (detectors 1 and 2), which we compare with the power spectrum of the non-deglitched pair-differenced TOD. We see in figure \ref{fig:TOD_psd_diff} that a simple 5 $\sigma$ deglitching method, which removes the majority of the direct hit signal, reduces the pair-differenced NEP by a factor of 10.

This conclusion implies that a hardware design goal of equal importance is to keep two detectors in one pair as geographically close to each other as possible such that the effective pair differencing occuring in the mapmaking procedure dominates the unpolarised signal. Hence, the ideal situation is one in which two detectors are kept in as much thermal contact with each other as possible. These findings have been further confirmed in \cite{tominaga2020simulation}.

If other mapmaking algorithms are chosen on the final LiteBIRD data, the level of contamination from common-mode cosmic ray noise is likely to change; a separate study will be the topic of future work.

\subsection{Scaling properties as a function of number of detectors}

Although we have simulated up to 64 detectors for the 119 GHz band on LFT, the wider implications of these results for LiteBIRD are found in the scaling factor of these results as a function of the number of detectors ($N_{\textrm{det}}$).

\begin{figure}[tbp]
\centering 
\includegraphics[width=0.9\textwidth]{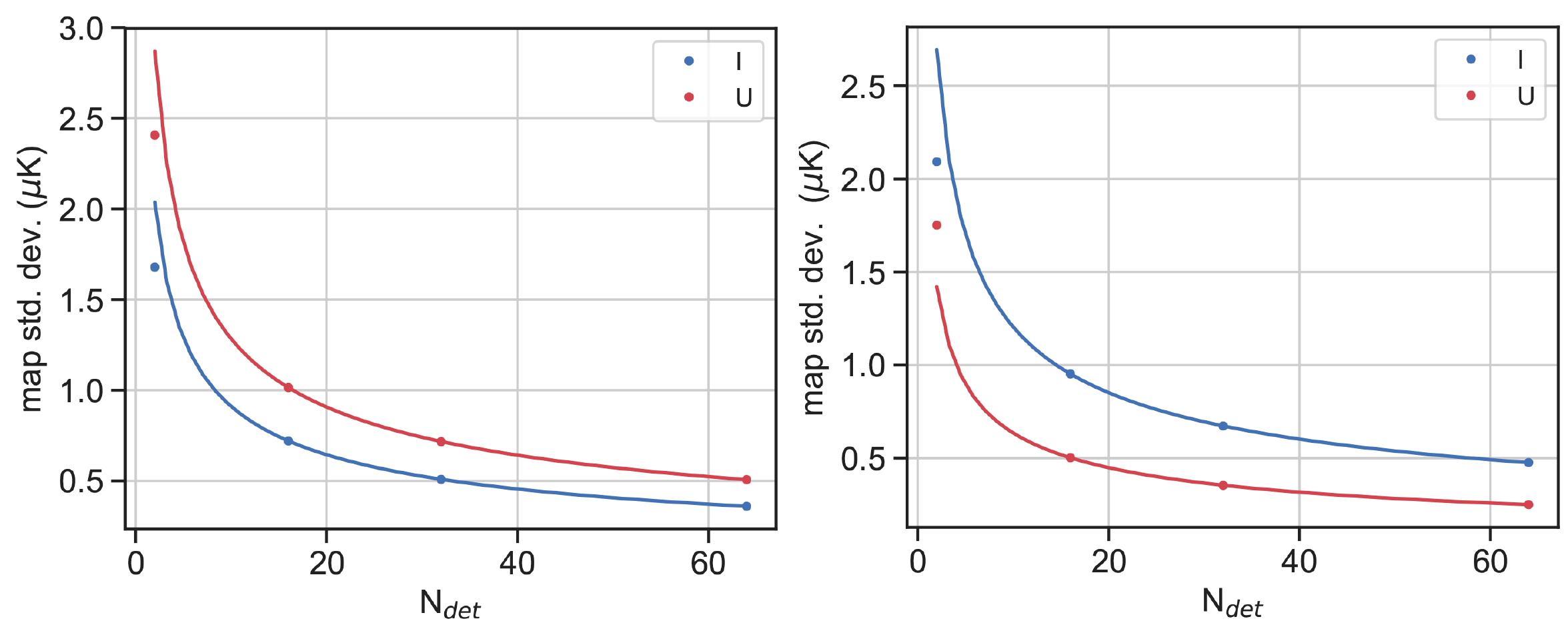}
\caption{\label{fig:detscale} \textit{Left:} The scaling of the map standard deviation as a function of $N_{\textrm{det}}$ without common-mode noise (left) and with common-mode noise (right).}
\end{figure}

We test this effect by taking the standard deviation of the simulated maps with and without the common-mode noise, finding that the common-mode case exhibits a higher variance in $I$ than in $U$, that this effect is inverted without the common-mode noise, and that in both cases, the map variance scales with $1/\sqrt{N_{\textrm{det}}}$.
These relationships are shown in figure~\ref{fig:detscale}.

With this in mind, it is important to note that the simulations we have presented are a worst-case scenario in many respects:
we have assumed a Planck HFI level of particle activity during the LiteBIRD observation time, we have simplified the interaction between incoming CRs and the spacecraft itself, and the wafer and detectors simulated do not contain any mitigation methods (although the development of detectors containing CR mitigation is currently under experimental investigation~\cite{westbrook2020detector,minami2020irradiation}). 

Furthermore, noting the decrease of $C_{\ell}^{BB}$ power from simulating a larger number of detectors, we expect the $C_{\ell}^{BB}$ power to further scale with the full number of detectors on LiteBIRD.
The \SI{119}{GHz} band alone contains 144 detectors on LFT and 488 detectors on MFT.
The specific scaling with respect to $N_{\mathrm{det}}$ is an effect which should be verified before full propagation of the noise source to degradation to the tensor-to-scalar ratio ($\delta r$).

The development of this end-to-end simulator framework is therefore an important and useful step towards probing the design and pipeline-side evaluation of the magnitude of the CR effect, but its immediate results should be taken as a proof-of-concept.
A more realistic estimate will be found through the successive sensitivity studies and optimisation planned, especially in relation to hardware changes and design assumptions.
The next section will describe one such example.

\subsection{Effect of assumed \texorpdfstring{$P_{\mathrm{opt}}$}{optical power} on simulation outcomes}
\label{sec:tomma2}
As described in section~\ref{sec:tomma}, the response of the TES to the cosmic ray noise assumes, in part, a background optical power of $P_{\mathrm{opt}}= \SI{0.5}{pW}$ coming from the CMB. From this assumption we fix the thermal conductance to $G=\SI{33.1}{pW/K}$.
Like many of the parameters chosen for our simulations, this is a conservative estimate. The actual estimated value of $P_{\mathrm{opt}}$ in the current LiteBIRD instrument model varies between \SI{0.29}{pW} at \SI{40}{GHz} to \SI{0.39}{pW} at \SI{166}{GHz}.
In order to test the effect of $P_{\mathrm{opt}}$ on the overall noise response of the detector, we have simulated the same CR events (in both temperature and power) whilst assuming $P_{\mathrm{opt}}$ between $0.1$ and \SI{1}{pW}, and hence $G$ between $7$ and $\SI{66.2}{pW/K}$.

\begin{figure}[tbp]
\centering 
\includegraphics[width=0.7\textwidth]{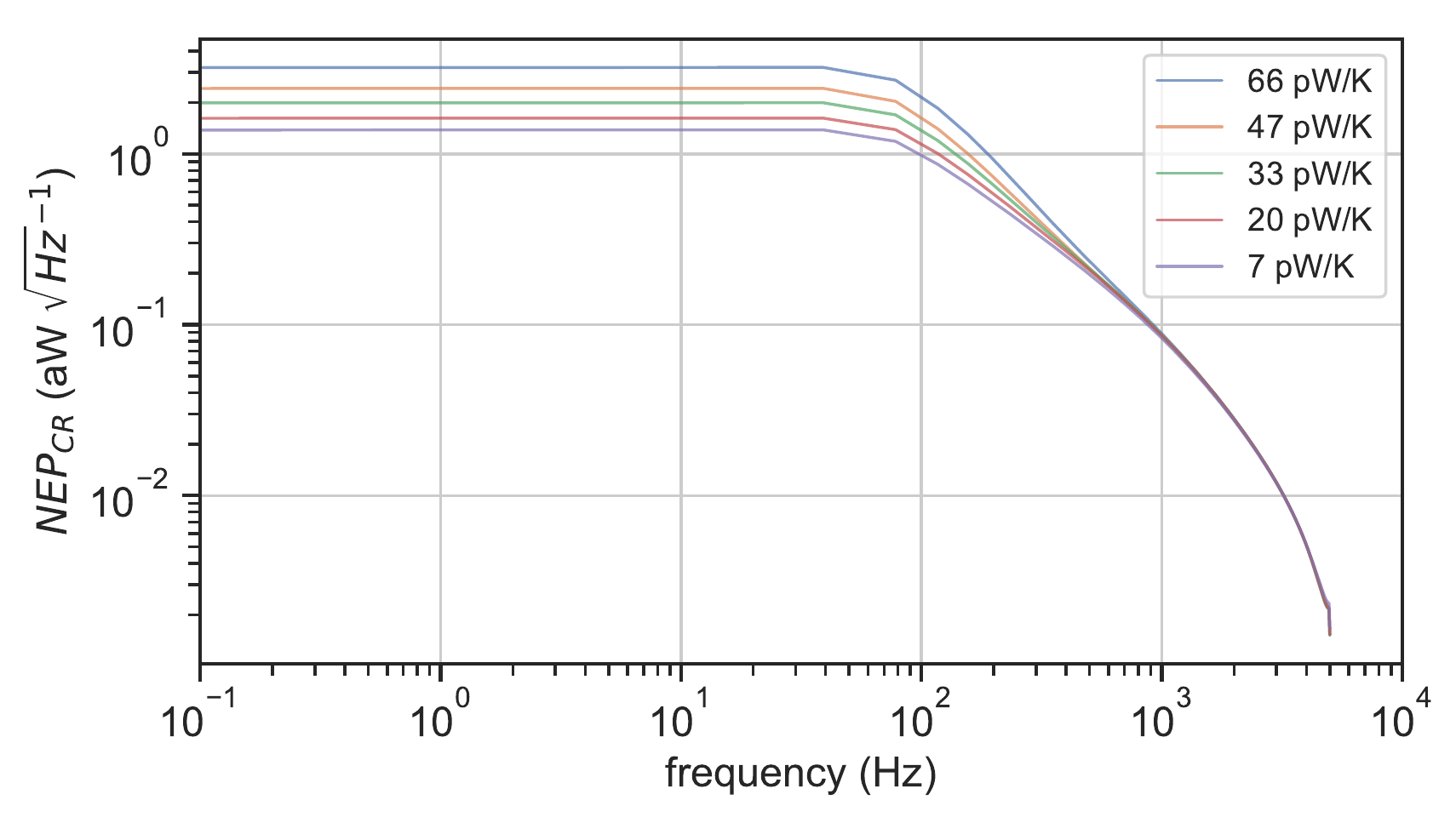}
\caption{\label{fig:Popt} Power spectral density of simulated TOD as a function of variable optical power assumed on the detector.}
\end{figure}

The power spectra arising from this study are shown in figure \ref{fig:Popt}. We find a variability of overall noise level (and thus NEP) as a function of $G$, scaling the NEP by a factor of $\approx{8}$ at the extrema of the study. This sensitivity study is therefore an example of the importance of follow-up investigation and careful assessment of all parameters assumed in these activities, because the outcome can be strongly affected by any changes to our current design.
We expect that future iterations of the end-to-end simulations will include fine-tuning of these parameters to be consistent with the LiteBIRD instrument model.

\subsection{Future work}

The first and most important task we plan to carry out is to investigate the addition of wirebonds to increase the $G$ of the detector wafer, and therefore the speed with which thermal fluctuations arising from the CR impacts in the wafer are evacuated to the thermal bath.
This will have an overall effect on the level of the thermal noise from indirect CR impacts, as has been found in other studies~\cite{stever2020thermal}.
Due to the large size of the LiteBIRD detector wafers, as well as their uncharacteristic thickness compared with other space missions, efficient thermal evacuation to the thermal bath is of crucial importance.
On the other hand, a wafer with too high of a thermal conductance will be more sensitive to individual CR hits.
This necessitates a study in which we derive the optimal $G$ of the wafer which takes this balance into account.

At present, the only thermal coupling between the thermal bath and the detector wafer is via the clamping force and the Invar holder.
If this clamping force is weak or variable across different wafers on the focal plane, the thermal evacuation and hence the overall level of common-mode noise will be inconsistent, as has been observed in other missions, e.g., OLIMPO~\cite{paiella2020flight}.
Steps for experimentally deducing the baseline $G$ of the clamping structure are being taken at UC Berkeley, and this will couple with forthcoming simulations to determine the subsequent dependence of thermal noise with the addition of wirebonds within the end-to-end simulator.

As mentioned in the previous section, we plan to further verify the scaling of the CR $C_{\ell}^{BB}$ as a function of $N_{\mathrm{det}}$ by altering the framework to simulate a larger number of detectors within one band, as well as to evaluate the variability across multiple wavebands.
With better-understood scaling laws, the level of $\delta r$ from CRs will be more confidently determined.

Finally, we plan to begin a number of sensitivity studies similar to the one already performed in section~\ref{sec:tomma2}.
This can include various changes to the hardware framework as described above, as well as the interaction between the data and the readout systems, and investigation into data postprocessing techniques (including, but not limited to, using machine learning techniques to find and subtract CRs in TOD).
These sensitivity studies will not only allow us to determine the real level of the CR threat, but also to develop a set of requirements that must be met in order to achieve full mission success.

\section{Conclusions}
We have presented an end-to-end simulator for evaluating the effect of cosmic ray systematic effects, and particularly the common-mode noise induced by this effect, on the mission outcomes of the LiteBIRD space mission. Cosmic ray systematic effects will continue to be an important consideration for any low-temperature space-borne mission, and a complete understanding and control of this systematic effect is vital to mission success.

Our current simulator couples the assumed radiative environment at L2, the thermal response of the LiteBIRD detector wafer, and the electrothermal response of the LiteBIRD detectors, generating time-ordered data of only predicted cosmic ray noise. The noise consists of a white noise component due to thermal fluctuations in the wafer, with a strong common-mode coupling across all detectors in a pixel, as well as indirect but high-amplitude direct CR hits into the TES.
The statistical attributes of 90 minutes of TOD (for 16 detectors in 2 pixels) have been analysed and used in a \texttt{TOAST-LiteBIRD} routine which injects CR noise based on these templates.
Using this, we can generate 3 years of TOD for all detectors, which is able to be expanded to a greater number of detectors or to other wavebands.

Using the most basic possible assumptions and worst-case scenarios for the prediction of the radiative environment at L2, the thermal attributes of the detector wafer, and with no software level post-processing, the unmitigated cosmic ray $C_{\ell}^{BB}$ power spectrum for one detector pair is flat and has an overall level of $\SI{10e-3}{\mu K_{\mathrm{CMB}}^{2}}$.
Increasing the number of detectors to 32 decreases $C_{\ell}^{BB}$ by a factor of $10$ ($\SI{10e-4}{\mu K_{\mathrm{CMB}}^{2})}$ with common-mode noise injection and and by a factor of $100$ ($\SI{10e-5}{\mu K_{\mathrm{CMB}}^{2}}$) without common-mode noise.

We find that the effective pair differencing occuring in the \texttt{libmadam} mapmaking procedure strongly affects the level to which CR noise arises in the polarised and unpolarised power spectra. From this, we conclude that the optimal design for minimising CR noise is one in which two orthogonal detectors (pairs) are kept in strong thermal contact with one another, and that simple techniques to remove direct hits (e.g. $5 \sigma$ filtering) strongly reduce the CR signal.

The evaluation of the scaling properties of the $C_{\ell}^{BB}$ component of the cosmic ray noise as a function of the number of detectors is future work, as well as the development of several parallel sensitivity studies and deglitching mechanisms.

This end-to-end simulator is an important tool for assessing the overall magnitude of this effect, and provides a vital framework for probing system design changes and development of a robust set of requirements necessary to achieve the successful measurement of cosmological $B$-modes.

\acknowledgments

This work was supported by JSPS KAKENHI Grant Numbers JP19K23438 and JP18K03715. 
GP acknowledges financial support from the APRA grant: ``Overcoming Systematic Effects in Cosmic Microwave Background Satellite Missions'', Grant number: 80NSSC19K0697. This research also used resources of the National Energy Research Scientiﬁc Computing Center (NERSC), a U.S. Department of Energy Oﬃce of Science User Facility operated under Contract No. DE-AC02-05CH11231.
Kavli IPMU is supported by World Premier International Research Center Initiative (WPI), MEXT, Japan.



%

\bibliographystyle{JHEP}
\bibliography{bib} 
\end{document}